\documentclass[aps,prb,showpacs,twocolumn,superscriptaddress,groupedaddress]{revtex4-1}

\usepackage{graphicx}  
\usepackage{dcolumn}   
\usepackage{bm}        
\usepackage{amssymb}   

\usepackage{rotating}

\usepackage{longtable}

\begin{document}

\title{Thermal vacancies in random alloys in the
single-site mean-field approximation}

\author{A.~V. Ruban}
\affiliation{Department of Materials Science and Engineering, KTH Royal
Institute of Technology, SE-100 44 Stockholm, Sweden}
\affiliation{Materials Center Leoben Forschung GmbH, A-8700 Leoben,
Austria}

\date{\today}
\begin{abstract}
A formalism for the vacancy formation energies in random alloys
within the single-site mean-filed approximation, where vacancy-vacancy
interaction is neglected, is outlined. It is shown that the alloy
configurational entropy can substantially reduce the concentration of
vacancies at high temperatures.
The energetics of vacancies in random Cu$_{0.5}$Ni$_{0.5}$ alloy
is considered as a numerical example illustrating the developed
formalism. It is shown that the effective formation energy is
increases with temperature, however, in this particular system
it is still below the mean value of the vacancy formation
energy which would correspond to the vacancy formation energy 
in a homogeneous model of a random alloy, such as given by
the coherent potential approximation.
\end{abstract}

\maketitle

Concentration of vacancies is one of the key parameters, which
determines the kinetic of phase transformation and diffusion
in solids. In spite of the structural simplicity of vacancies,
their energetics has proven to be one of the least reliable
physical properties determined in the first principles calculations
(see, for instance, Ref. 
\onlinecite{am05-1,am05-2,pbesol,mattsson08,nazarov12,glensk14}).
The situation becomes even more complicated at high temperatures,
where anharmonic effects play an important role.\cite{glensk14}

In this paper, we will not however deal with those problems
related to different approximations in first-principles
calculations and subsequent modelling of the vacancy thermodynamics,
but rather consider another important aspect, namely, the
statistical description of vacancies in concentrated alloys
at finite temperature connected with their first-principles
modelling. This topic has recently been recently attracted
attention of several groups doing first-principles
simulations.\cite{ven05,muzyk11,piochaud14,zhang15,belak15}
In contrast to those investigations, in this work a simplified
model for the energetics of vacancies will be presented for
completely random alloys with the purpose to get a qualitative
picture of the configurational effects.

It is based on the single-site mean-field approximation, and thus 
all the effects related to the vacancy-vacancy interactions
will be ignored, while vacancy-alloy-component interactions
will be indirectly taken into consideration through the account
of the local environment effects next to the vacancy.
Although this is a simplified model, it anyway yields a quite
accurate description of the phenomenon in real systems.
To demonstrate the formalism, we will consider the energetics of
vacancies in Cu$_{0.5}$Ni$_{0.5}$ random alloy.

The vacancy formation energy at 0 K in a binary random
A$_{c}$B$_{1-c}$ alloy can be formally defined as

\begin{equation} \label{eq:E_f}
E_f^0 =
 \min\frac{d E_0(\rm A_{\it c(1 - c_v)} \rm B_{\it (1-c)(1-c_v)}
\rm Va_{\it c_v})}{d c_v} \mid_{c_v = 0}  ,
\end{equation}
where E$_0$ is the total energy per atom of a random
A$_{c (1 - c_v)}$B$_{(1-c)(1-c_v)}$Va$_{c_v}$ alloy consisting $c_v$ 
concentration of vacancies (Va). This definition takes into
consideration the fact that the derivative in (\ref{eq:E_f})
is not well defined since in real random
alloys there exist substantial fluctuations of local compositions,
which affect this derivative leading to a wide spectrum of the
{\em local vacancy formation energies} connected to the specific
space arrangements of the alloy components around the vacancy.
At 0 K, the vacancy formation energy, $E_f^0$ is apparently
determined by the lowest value of the derivative in
(\ref{eq:E_f}). Definition (\ref{eq:E_f}) also formally implies
that the ratio of the concentrations of A and B
alloy components is not change during vacancy formation.

The dependence of the vacancy formation energy on the local
environment can be also viewed as interaction energy between
vacancy and alloy components. Nowadays,
it can be obtained in first-principles calculations using,
for instance, the so-called "local cluster expansion".\cite{ven05,zhang15}
If a supercell approach is used to determine local vacancy formation
energies in random alloys, these effects can be naturally
reproduced since the fluctuations of the local environment
around each site are inevitable.

The existence of the local environment effects becomes important
at finite temperatures, where vacancies with higher formation
energies can be also created. For a given alloy configuration 
one can introduce the local vacancy formation
energy distribution function, $g(E)$, which determines the number
of sites, $Ng(E)$ in the alloy sample of size $N$, where the
local vacancy formation energy is $E$, which satisfies the
following normalisation:

\begin{equation}
\int dE g(E) =  1 .
\end{equation}

At finite temperatures, $g(E)$ determines the distribution of
vacancies with respect to their local environment.
To obtain it, we first define {\it effective} vacancy formation
energy or free energy, which connects the free energy of the system
with concentration of vacancies in a phenomenological way.
For a binary random A$_c$B$_{1-c}$ alloy, it is defined as

\begin{equation} \label{eq:G_vac}
G_{\rm vac} = c_v \bar{G}_f - T S_{conf} ,
\end{equation} 
where $c_v$ is the equilibrium concentration of vacancies;
$\bar{G}_f$ is the effective vacancy formation free energy
and $S_{conf}$ the configurational entropy of an alloy with
vacancies:

\begin{eqnarray} \label{eq:S}
S_{conf} &=& -[ c_v \ln c_v +  \\ \nonumber
 && c_A \ln c_A + c_B \ln c_B ],
\end{eqnarray}
where $c_A = c(1 - c_v)$ and $c_B = (1 - c)(1 - c_v)$ are the
concentration of alloy components, which implies that the ratio
of concentrations of both components remains constant and the
same as in the alloy without vacancies.

In the single-site approximation, the minimisation of (\ref{eq:G_vac})
with respect to $c_v$
under the condition that the concentration of vacancies is
substantially smaller than that of alloy components yields:

\begin{equation} \label{eq:c_v}
c_v = \exp \left[-\frac{\bar{G}_f + TS_{all}}{T}\right] \equiv 
\exp \left[-\frac{\widetilde{G}_f}{T}\right] ,
\end{equation}
where $S_{all} = - [c \ln c + (1-c) \ln (1-c)]$ is the alloy
configurational entropy without vacancies and
$\widetilde{G}_f = \bar{G}_f + TS_{all}$
is the renormalised vacancy formation energy due to the
randomness of the alloy.

This result shows that the alloy configurational entropy can
substantially reduce the concentration of vacancies in alloys.
For instance, in the equiatomic binary random alloy (c=0.5),
the equilibrium concentration is reduced by a factor of 2
compared to that in pure metal.
At 1500 K, it corresponds to an approximate increase of
the effective vacancy formation energy of about 0.09 eV.
Let us note that the above derivation holds for multicomponent
alloys, where this effect can be much more pronounced.
For instance, in a four-component equimolar (frequently called
"high entropy") random alloy the concentration of 
vacancies will be 4 times lower than that in pure metal
having the same vacancy formation energy, which corresponds to
the additional increase of the effective vacancy formation
energy of about 0.18 eV at 1500 K.

Considering vacancies at different sites as independent, i.e.
neglecting vacancy-vacancy interaction and assuming that the vacancy
formation entropy, $S_f$, associated with vibrational, magnetic and
electronic degrees of freedom, does not depend on the local environment,
it is easy to show that

\begin{equation} \label{eq:c_v_g}
c_v = \exp(S_f - S_{all})\int dE g(E) \exp\left(-\frac{E}{T} \right) .
\end{equation}
Otherwise one should consider the distribution function for the
local vacancy formation free energies, $g_G(G)$. The expression under
the integral in (\ref{eq:c_v_g}) is just the concentration of vacancies
for specific energy formation
$E$: $c_v(E) =  g(E) \exp\left(-\frac{E}{T} \right)$.
Comparing (\ref{eq:c_v}) and (\ref{eq:c_v_g}), one finds that

\begin{equation}
\bar{G}_f = -T \ln \left[ \int dE g(E) \exp\left(-\frac{E}{T}\right)\right] -
T S_f ,
\end{equation} 
or the effective vacancy formation energy, $\bar{E}_f$ is

\begin{equation} \label{eq:E_eff}
\bar{E}_f = -T \ln \left[ \int dE g(E) \exp\left(-\frac{E}{T}\right)\right] ,
\end{equation}
while the renormalised vacancy formation energies will have an additional
contribution $TS_{all}$: $\widetilde{G}_f = \bar{G}_f + TS_{all}$
and $\widetilde{E}_f = \bar{E}_f + TS_{all}$. 

Let us now consider vacancy energetics in Cu$_{0.5}$Ni$_{0.5}$ random
alloy. It should be stressed again that only a configurational part of
the problem will be considered here, without any complications
related to other thermal effects, such as electronic, vibrational
or magnetic excitations. We therefore also disregard thermal
lattice expansion and perform calculations for a fixed lattice
parameter of 3.56 \AA. 

To determine the local vacancy formation energies, we use the
exact-muffin-tin orbital locally self-consistent Green's function
(ELSGF) method,\cite{peil12}
which allows relatively accurate first-principles calculations
of the vacancy formation energies, at least on a rigid lattice
without a consideration of the local lattice relaxations. The
latter may decrease the vacancy formation energy by 0.1 -- 0.2 eV,
which is comparable with the usual error due to the use of 
different exchange-correlation approximations.
The supercell size has been chosen to be 108 atoms
(a 3$\times$3$\times$3 cell build upon the 4-atom cubic fcc cell).

Every atom in this supercell was exchanged by a vacancy, 
and then
the local vacancy formation energy at site $i$, $E_f^i$, has
been determined as

\begin{equation}
E_{f}^i = E_{vac}^i - \frac{N-1}{N} E_{all} -
(N-1) \Delta c \mu_{eff} ,
\end{equation} 
where $E^i_{vac}$ is the total energy of the supercell with vacancy
at site $i$; $E_{all}$ the total energy of the defect free supercell;
$N$ is the number of atoms in the supercell; $\Delta c$ is the change
of the supercell composition due to vacancy formation (for instance,
in our case $\Delta c = \pm(53/107 - 54/108)$, and
$\mu_{eff}$ is the effective chemical potential of the alloy
determined as

\begin{equation}
\mu_{eff} = \frac{\partial E_0(\rm A_c \rm B_{1-c})}{\partial c} .
\end{equation}
Here, the $E_0$ is the total energy per atom of random A$_c$B$_{1-c}$
alloy. The latter can be quite accurately (and what is important: 
consistently with the LSGF calculations) obtained by the EMTO-CPA
method\cite{vitos01,vitos07} using the Lyngby version of the
code\cite{code} with the appropriate choice of
the electrostatic screening
constants (determined again from the corresponding ELSGF
supercell calculations\cite{screening}). 

Other details of the calculations are the following.
The partial waves up to $l_{max}=3$ were
used in the self-consistent calculations. The total energies
have been obtained using the full charge density technique.\cite{vitos07}
The ELSGF
calculations have been performed using the local interaction
zone (LIZ) which included the first two coordination shells around
the central site. This
means that chemical configurational effects were effectively
cut off beyond the second coordination shell (which is not the
case of electrostatic interactions, although they are relatively
weak in this system, and some multisite interactions for the
clusters within the LIZ). The PBE-sol exchange-correlation
potential\cite{pbesol} has been used, which is partly the reason
for the difference of the present results and those of
Ref. \onlinecite{zhang15}.

\begin{figure}[tb]
\centering
\includegraphics[width=0.9\linewidth]{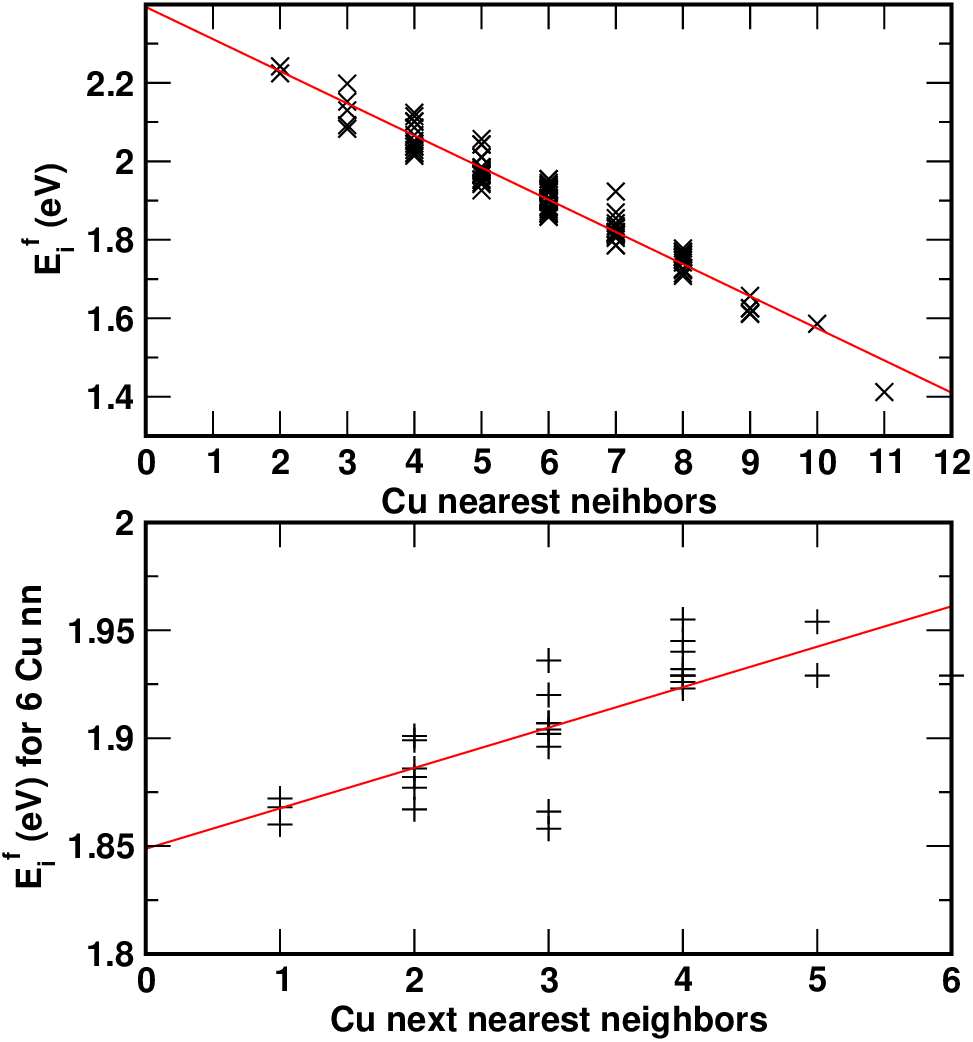}
\caption{(Color online) 
Local vacancy formation energies in 108-atom supercell
representing a random Cu$_{0.5}$Ni$_{0.5}$ alloy. The
distribution of the local vacancy formation energies with
respect to the number of the Cu nearest neighbours is shown in
the top panel of the figure, while the distribution with
respect to the number of the next nearest neighbours is shown in
lower panel. In the latter case only sites having 6 Cu nearest
neighbours are included in the figure. Straight lines show the
average slopes, which corresponds to the vacancy-Cu interaction
at the first and second coordination shell, respectively.}
\label{fig:E_loc}
\end{figure} 

In Fig. \ref{fig:E_loc}, the local vacancy formation energies are
shown as a function of the number of Cu atoms next to the
vacancy.\cite{comp} Although there is a dispersion of the local vacancy
formation energies for every number of Cu nearest neighbours,
they almost linearly decrease with the number of Cu nearest neighbours.
The slope of the average descent of the local energies is
in fact the vacancy-Cu (or vacancy-Ni if taken with the
opposite sign) interaction energy, which is approximately
$-$0.082 eV for the first and 0.018 eV for the second 
coordination shells. The dispersion is due to other type
of interactions.

It should be mentioned that there is no apparent
dependence of the local vacancy formation energies on the
type of the atom occupying this site in the defect
free supercell. This contrasts with the results obtained 
in Ref. \onlinecite{zhang15} where much smaller supercells
have been used.
From a general point of view, such a dependence should not
exist in the macroscopic limit, unless a ghost of the 
removed atom is still in the site. Although in the reality
nobody is certain about ghosts, it cannot exist in the well
determined first-principles calculations.

The spurious dependence can originate from some technical details
of the modeling. For instance, it is clear that
small supercells, of an order of tens of atoms, provide quite
a bad model for investigation of the local environment effects
due to the fact that no good statistics can be obtained
just from several sites. Besides, every exchange of an
atom by vacancy leads to the different (from the initial)
on average atomic distribution correlations functions. 

The difference in statistics of the local environment for
different alloy components of course also exists in the case of
the used here 108-atom supercell, where
the representation of the possible local environment effects
is also quite restricted. It can be clearly seen in Fig. \ref{fig:E_loc}
that there are no sites in the supercell completely surrounded
by Cu or Ni atoms, and there is only one site with 11 Cu nearest
neighbours, while there are no sites with 11 Ni nearest neighbours.

\begin{figure}[tb]
\centering
\includegraphics[width=0.9\linewidth]{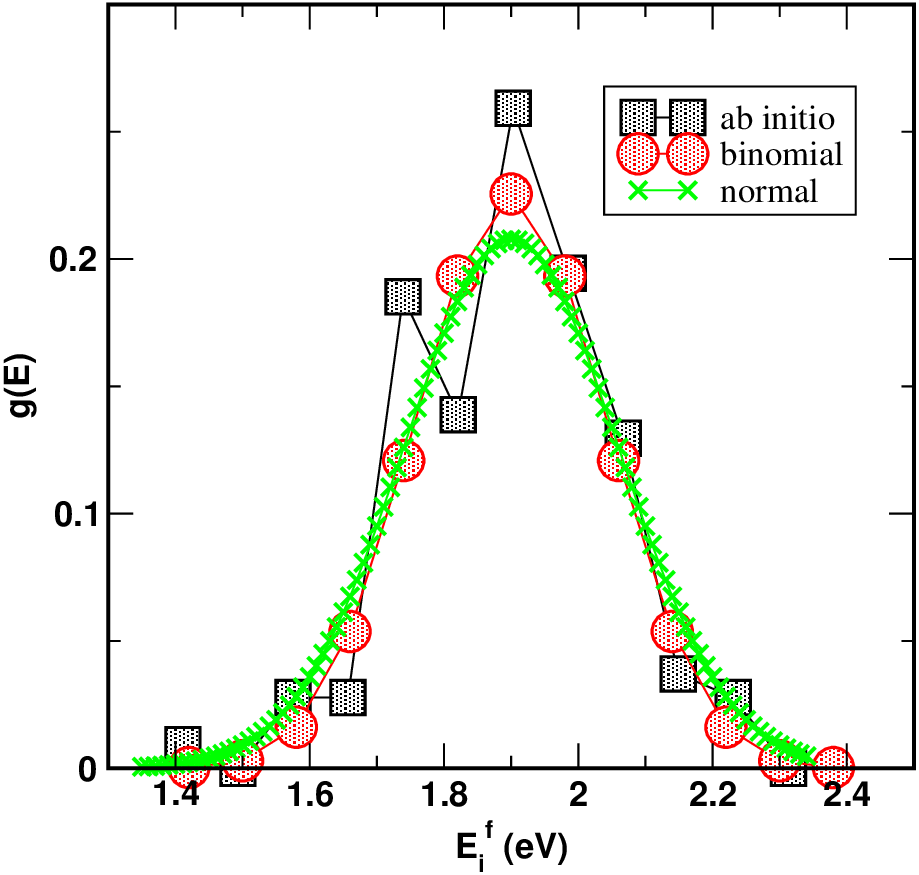}
\caption{(Color online) 
Local vacancy formation energy distribution functions:
squares are the results of the 108-atom supercell
calculations;
circles are binomial distribution (see text) and
crosses are normal distribution.}
\label{fig:gE}
\end{figure} 

In spite of this fact, one can still establish a qualitatively 
clear picture of the local environment effects in alloy.
In Fig. \ref{fig:gE}, the local vacancy formation energy distribution
function, $g(E)$, obtained from the present {\em ab initio}
calculations is shown. It was calculated using 0.08 eV energy
interval window, which corresponds to the average change of the
local vacancy formation energy when the number of the Cu
nearest neighbours changes by one. As one can see, it can be
very well approximated by the discreet binomial distribution,
which for a binary equiatomic alloy is

\begin{equation}
g_b(E(n)) = \frac{12!}{2^{12} n!(12-n)!}  ,
\end{equation}
for $n$ going from 0 to 12 and $E_f(n) = E_f^0 + n V_1$ where
$E_f^0$ is the lowest local vacancy formation energy (as it is 
determined in Eq. (\ref{eq:E_f})),
which corresponds to the case $n=0$ and $V_1$ is the 
{\em positive} interaction
energy between the vacancy and the counted by $n$ alloy component.
It is clear that such a choice of interaction, which is positive
in this case, can be always made. In our case, it corresponds
to the vacancy-Ni interaction and thus $n$ is the number of Ni atoms
next to the vacancy.

Equally, the local vacancy formation energy distribution
function, $g(E)$, can be approximated by the continues normal
distribution (for the equiatomic composition only) as

\begin{equation}
g_n(E) = \frac{1}{\sigma \sqrt{2 \pi}}
\exp\left[-\frac{(E - \langle E_f \rangle)^2}{2 \sigma^2}\right] ,
\end{equation}
where $\langle E_f \rangle$ is the mean local vacancy formation energy,
which is about 1.9 eV in this particular case, and
$\sigma = 2 |V_1|$. 

\begin{figure}[tb]
\centering
\includegraphics[width=0.9\linewidth]{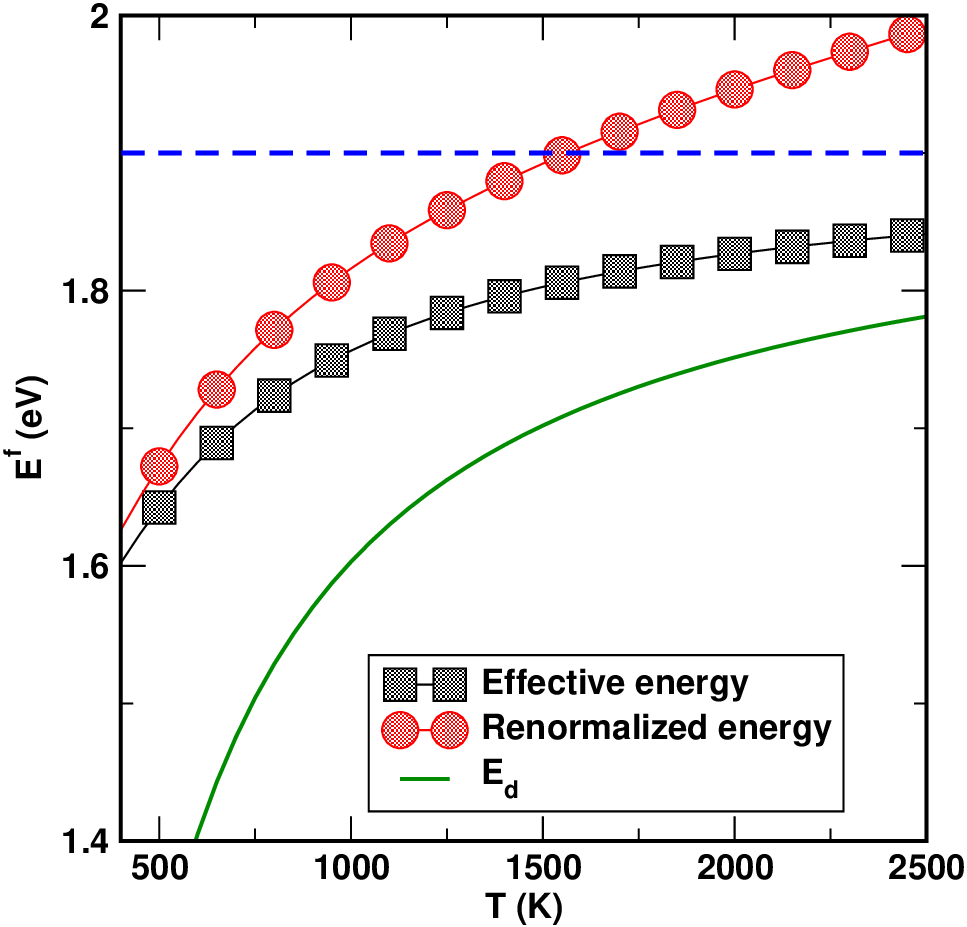}
\caption{(Color online) 
Effective ($\bar{E}_f$) and renormalised ($\widetilde{E}_f$) vacancy
formation energies in random Cu$_{0.5}$Ni$_{0.5}$ alloy obtained
as a function of temperature neglecting all the possible type
of thermal excitations except configurational in the single-site
mean field approximation. 
The dashed line shows the mean value
of the vacancy formation energy, $\langle E_f \rangle$, which one,
for instance, would obtain in the homogeneous CPA calculations.}
\label{fig:E_eff}
\end{figure}

Using $g_n(E)$ and (\ref{eq:E_eff}), one can calculate the
effective, $\bar{E}_f$, and
renormalised, $\widetilde{E}_f$, vacancy formation energies as
functions of temperature 
(no thermal lattice expansion and other effects are included). They
are shown in Fig. \ref{fig:E_eff}. As one can see, both vacancy 
formation energies, effective and renormalised, exhibit quite
strong dependence on the temperature at low temperatures, while
at higher temperatures, $\bar{E}_f$ changes quite little and
$\widetilde{E}_f$ grows linearly with temperature.
It is interesting that at least in this particular case
$\bar{E}_f$ does not reach the mean value, $\langle E_f \rangle$ even
at relatively high temperatures.

In fact,  $\langle E_f \rangle$ corresponds to the vacancy
formation energy obtained in the homogeneous CPA calculations
like those in Ref.
\onlinecite{delczeg12,delczeg15},
where all the sites of the supercell are treated as effective
CPA medium of the given alloy composition.\cite{vac_CPA} This means, that such
energies do not make much sense in systems, where the local vacancy
formation energies strongly
depend on their local environment, like Cu-Ni calculated here.

Another energy of interest, which we call here $E_d$, is the local
vacancy formation energy, which yields dominating contribution to
the vacancy concentration at a given temperature. It is related to
the dominating type of the local environment of vacancies at given T
and can be found by maximising $c_v(E)$.
In the case of a binary equiatomic alloy, it can be approximately
obtained using the normal distribution  $g_n(E)$ of the local
vacancy formation energies:

\begin{equation} \label{eq:E_d}
E_{d} = \langle E_f \rangle -\frac{\sigma^2}{T} = 
\langle E_f \rangle - \frac{4 V_1^2}{T} .
\end{equation}  
It is shown in Fig. \ref{fig:E_eff}. As one can see, it is
less than the effective formation energy, although at low
temperatures, its definition (\ref{eq:E_d}) breaks down
since $g_n(E)$ is always non-zero
for all positive energies, while $g(E)$ of a real system
is non-zero only within some specific energy interval
above $E_f^0$.

Now, we can estimate the preferential local environment of
vacancies at a given temperature. 
Since $\langle E_f \rangle \approx E_f^0 + (z_1/2) V_1$,
where $z_1$ is the number of the nearest
neighbour sites, the number of Ni atoms next to the vacancy
with the local formation energy $E_{d}$ at temperature $T$ is 

\begin{equation} \label{eq:n_E}
n_{\rm Ni}(E_{d}) = \frac{z_1}{2} - \frac{4V_1}{T} .
\end{equation}

This is a quite interesting result showing first of all
that this number is inverse proportional to the temperature
and, secondly, it is always less than $z_1/2$, which is just the
average number of Ni atoms of the equiatomic random alloy
considered here, reaching it is maximum, $z_1/2 = c_{\rm Ni}$,
only at infinite temperature. This again shows that a homogeneous
CPA-like model of vacancies in random alloys corresponds this 
infinite temperature limit and thus should always overestimate the
vacancy formation energy if there is non-negligible
vacancy-alloy-component interaction.

It is obvious, that the number of Cu atoms next to the vacancy
with the local formation energy $E_{d}$ at temperature $T$ is
$n_{\rm Cu}(E_{d}) = z_1/2 + 4V_1/T$ or in 
general in high temperature limit
$n_{\rm Cu}(E_{d}) = z_1 c_{\rm Cu} + 4V_1/T$.\cite{T_limit}
This kind of asymptotic behaviour is observed for the {\em average
number} of Cu nearest neighbours next to the vacancy as a function
of temperature in Ref. \onlinecite{zhang15} presented in Fig. 12,
where one can clearly see inverse temperature dependence of
this number on the temperature and the fact that the minimal average
number of Cu atoms next to the vacancy
in the limit $T \to \infty$, is 3 and which is the average
number of Cu atoms in random Cu$_{0.25}$Ni$_{0.75}$ alloy
at the first coordination shell ($z_1 c_{\rm Cu}$).

In summary, a single-site mean field theory for thermal vacancies
in random alloys is presented. It shows that the alloy configurational
entropy renormalises the effective vacancy formation energy, and this
contribution linearly increases with temperature. As a numerical example,
we have calculated the vacancy formation energies in Cu$_{0.5}$Ni$_{0.5}$
random alloy and demonstrated that configurational effects play important
role. In particular, the effective formation energy is lower than the
mean value of the local vacancy formation energy, and this effect is
proportional to the vacancy-solute/solvent interactions.

\acknowledgments

This work has been initiated after some discussion of the
vacancy formation energies in random alloys with Hu-Bin Luo
from the Key Laboratory of Magnetic Materials and Devices, Ningbo
Institute of Material Technology and Engineering, Chinese
Academy of Sciences. The author
acknowledges the support of the Swedish Research Council
(VR project 15339-91505-33), the European Research Council grant,
the VINNEX center Hero-m, financed by the  Swedish Governmental
Agency for Innovation Systems (VINNOVA), Swedish industry, and
the Royal Institute of Technology (KTH). Calculations have been
done using NSC (Link\"oping) and PDC (KTH) resources provided
by the Swedish National Infrastructure for Computing (SNIC).


\begin{thebibliography}{0}

\bibitem{am05-1} R. Armiento, A.~E. Mattsson, Phys. Rev. B {\bf 72},
085108 (2005)

\bibitem{am05-2} A.~E. Mattsson, R.~R. Wixom, R. Armiento,
Phys. Rev. B {\bf 77}, 155211 (2008). 

\bibitem{pbesol} J. P. Perdew, A. Ruzsinszky, G.~I. Csonka,
O.~A. Vydrov, G.~E. Scuseria, L.~A. Constantin, X. Zhou, K. Burke,  
Phys. Rev. Lett. {\bf 100}, 136406 (2008).

\bibitem{mattsson08} A.~E. Mattsson, R. Armiento, J. Paier,
G. Kresse, J.~M. Wills, T.~R. Mattsson, J. Chem. Phys. {\bf 128},
084714 (2008).

\bibitem{nazarov12} R. Nazarov, T. Hickel, J. Neugebauer,
Phys. Rev. B {\bf 85}, 144118 (2012).

\bibitem{glensk14} A. Glensk, B. Grabowski, T. Hickel, and
J. Neugebauer, Phys. Rev. X {\bf 4}, 011018 (2014).

\bibitem{ven05} A. Van der Ven and G. Ceder, Phys. Rev. B
{\bf 71}, 054102 (2005).

\bibitem{muzyk11} M. Muzyk, D. Nguyen-Manh, K. J. Kurzydlowski,
N. L. Baluc, and S. L. Dudarev, Phys. Rev. B {\bf 84}, 104115
(2011).

\bibitem{piochaud14} J.~B. Piochaud, T.~P.~C. Klaver, G. Adjanor,
P. Olsson, C. Domain, and C.~S. Becquart, Phys. Rev. B {\bf 89},
024101 (2014).

\bibitem{zhang15} X. Zhang and M. H. F. Sluiter, Phys. Rev. B
{\bf 91}, 174107 (2015).

\bibitem{belak15} A.~A. Belak and A. Van der Ven, Phys. Rev. B {\bf 91},
224109 (2015).

\bibitem{peil12} O.E. Peil, A.V. Ruban, and B. Johansson,
Phys. Rev. B {\bf 85}, 165140 (2012).

\bibitem{vitos01} L.~Vitos, I.~A.~Abrikosov, and B.~Johansson,
Phys.~Rev.~Lett. \textbf{87}, 156401 (2001).

\bibitem{vitos07} L. Vitos, \textit{Computational Quantum Mechanics for
Materials Engineers} (Springer-Verlag, London, 2007).

\bibitem{code} The Lyngby version of the EMTO code properly takes
into consideration electrostatics in random alloys in contrast to
other existing versions. It is distributed by the author of the
paper.

\bibitem{screening} A.~V. Ruban and H.~L. Skriver, Phys. Rev. B
{\bf 66}, 024201 (2002); A.~V. Ruban, S.~I. Simak, P.~A. Korzhavyi,
and H.~L. Skriver, ibid, 024202 (2002).

\bibitem{comp} The local energies are approximately about 0.3-0.4 
eV below the results of Ref. \onlinecite{zhang15} due to 
different exchange correlation potential used in the calculations
(see, for instance Table III in Ref. \onlinecite{zhang15}) and
neglect of local lattice relaxation effects in this work.

\bibitem{delczeg12} L. Delczeg, B. Johansson, and L. Vitos,
Phys. Rev. B {\bf 85}, 174101 (2012).

\bibitem{delczeg15} E. K. Delczeg-Czirjak, L. Delczeg, L. Vitos,
and O. Eriksson
Phys. Rev. B 92, 224107 (2015).

\bibitem{vac_CPA} The EMTO-CPA calculations of the vacancy formation
energy using a 32-atom supercell (2$\times$2$\times$2($\times$4)) where
all the atomic positions are "occupied" by the inhomogeneous CPA
effective medium of Cu$_{50}$Ni$_{50}$ random alloy yields the
vacancy formation energy 1.87 eV, which is in good agreement with
the ESLGF mean value of 1.9 eV taking into consideration that that
these are two completely different techniques, with different details
of calculations such as the size of the supercell, k-point mesh,
CPA small inaccuracies and so on.

\bibitem{T_limit} Let us note that this simple analysis is valid 
only at sufficiently high temperatures due to the use of the
normal distribution, $g_n(E)$, in the derivation. The point is that
$g_n(E)$ is not restricted by a system specific energy interval, like
$g(E)$. Since $n(E_{d}) \geq 0$, the above consideration
makes sense only for $T \geq (8/z_1) V_1$. In our case it
is about 600 K. 
  


\end{thebibliography}
\end{document}